\documentclass[12pt]{article}

\usepackage{graphicx}

\newcommand{\ud}{\mathrm{d}}
\newcommand{\dyn}{\mathrm{dyn}}

\begin{document}
\begin{center}
{\Large\bf \boldmath Impact of eight-quark interactions 
       in chiral phase transitions I: Secondary magnetic 
       catalysis} 

\vspace*{6mm}
{A. A. Osipov$^{a,b}$, B. Hiller$^b$, A. H. Blin$^b$ and J. da 
 Provid\^encia$^b$}\\      
{\small \it $^a$ Dzhelepov Laboratory of Nuclear Problems, JINR, 141980 
                 Dubna, Russia\\  
            $^b$ Departamento de F\'{\i}sica, Universidade de Coimbra, 
                 P-3004-516 Coimbra, Portugal}
\end{center}

\vspace*{6mm}

\begin{abstract}
The influence of a constant magnetic f\mbox{}ield on the order 
parameter of the four-dimensional Nambu and Jona-Lasinio model 
extended by the 't Hooft six-quark term and eight-quark interactions 
is considered. It is shown that the multi-quark interactions cause 
the order parameter to increase sharply (secondary magnetic catalysis) 
with increasing strength of the f\mbox{}ield at the characteristic 
scale $H\sim 10^{14}\Lambda^2$ G/MeV$^2$. 
\end{abstract}

\vspace*{6mm}

It has been shown in a series of papers \cite{Lemmer:1989}-\cite{
Krive:1991} that in $2+1$ and $3+1$ dimensions a constant magnetic 
f\mbox{}ield $H\neq 0$ catalyzes dynamical symmetry breaking leading 
to a fermion mass even at the weakest attractive four-fermion interaction 
between massless particles, and the symmetry is not restored at any 
arbitrarily large $H$.
 
It is known, however, that the QCD motivated ef\mbox{}fective lagrangian 
for the light quarks $(N_f=3)$ contains also the six-fermion term: the 
$U(1)_A$ breaking 't Hooft interaction, and probably eight-quark terms. 
These extensions of the Nambu and Jona-Lasinio model are well-known, 
for instance, the four-quark $U(3)_L\times U(3)_R$ chiral symmetric 
lagrangian together with the 't Hooft six-quark interactions has been 
extensively studied at the mean-f\mbox{}ield level 
\cite{Bernard:1988}-\cite{Hatsuda:1994}. Recently it has been also shown 
\cite{Osipov:2006, Osipov:2007} that the eight-quark interactions are of 
vital importance to stabilize the multi-quark vacuum.

The additional multi-quark forces can af\mbox{}fect the result which is 
obtained when only four-fermion interactions are considered. We argue 
here that the 't Hooft and eight-quark interactions can modify the theory 
in such a way that the local minimum, catalyzed by the constant magnetic 
f\mbox{}ield, is smoothed out by increasing the strength of the 
f\mbox{}ield. This is an alternative regime to the known one in which the 
strong magnetic f\mbox{}ield cannot change the ground state of the system. 
For the f\mbox{}irst scenario to become possible it is 
suf\mbox{}f\mbox{}icient that the couplings of multi-quark interactions 
are chosen such that the system displays more than one solution of the 
gap equation at $H=0$. However, the above condition is not a requirement. 
Even if the gap equation has only one nontrivial solution at small $H$, 
an increase in the magnetic f\mbox{}ield can induce the formation of a 
second minimum. Starting from some critical value $H_c$ the second 
minimum is becoming a new ground state. We call this phenomenon a 
secondary magnetic catalysis.
To see the details we need the ef\mbox{}fective potential of the theory, 
$V(m,|QH|)=V_{st}+V_{S}$, which is the sum of two terms. The f\mbox{}irst 
contribution results from the many-fermion vertices, after reducing them 
to a bilinear form with help of bosonic auxiliary f\mbox{}ields, and 
subsequent integration over these f\mbox{}ields, using the stationary 
phase method. This part does not depend on the magnetic f\mbox{}ield.
The specif\mbox{}ic details of these calculations are given in our 
recent work \cite{Osipov:2006}. In the $SU(3)_f$ symmetric case the 
result is
\begin{equation} 
\label{effpot1}
     V_{st} = \frac{1}{16}
     \left(12Gh^2 + \kappa h^3 + \frac{27}{2}\lambda h^4\right). 
\end{equation}
The function $h$ is a solution of the stationary phase equation  
$12\lambda h^3 + \kappa h^2 +16(Gh+m)=0$, where $G,\kappa,\lambda$ are 
couplings of four, six and eight-quark interactions correspondingly.  
This cubic equation has one real root, if $G/\lambda > (\kappa /24
\lambda )^2$. Assuming that the couplings fulf\mbox{}ill the inequality, 
one f\mbox{}inds the single valued function $h(m,G,\kappa,\lambda )$. 

The second term, $V_S$, derives from the integration over the quark 
bilinears in the functional integral of the theory in presence of a 
constant magnetic f\mbox{}ield $H$. As has been calculated by Schwinger 
a long time ago \cite{Schwinger:1951} $V_{S}=\sum_{i=u,d,s}V_S(m_i,|Q_iH|
),$ where
\begin{equation}
   V_S(m, |QH|)=\frac{N_c}{8\pi^2}\int\limits_0^\infty 
   \frac{\ud s}{s^2} e^{-sm^2}\rho (s,\Lambda^2) 
   |QH|\coth (s|QH|)+\mathrm{const}.  
\label{SR}
\end{equation}
Here the cutof\mbox{}f $\Lambda$ has been introduced by subtracting 
of\mbox{}f suitable counterterms to regularize the integral at the
lower limit: $\rho (s,\Lambda^2)=1-(1+s\Lambda^2)e^{-s\Lambda^2}$. The 
unessential constant is chosen to have $V_S(0,|QH|)=0$. We ignore in 
the remaining the charge dif\mbox{}ference of $u$ and $d,s$ quarks: 
the averaged common charge $|Q|=|4e/9|$ will be used. 

One sees that the gap equation, $\ud V(m)/ \ud m = 0$, has always a 
trivial solution $m=0$, which corresponds to the point where the
potential reaches its local maximum, if $H\neq 0$. This phenomenon is 
known as magnetic catalysis of dynamical chiral symmetry 
breaking. The nontrivial solution is contained in the equation  
\begin{equation}
\label{gap}
   -\frac{2\pi^2h(m)}{\Lambda^2N_cm}=
   \psi\left(\frac{\Lambda^2+m^2}{2|QH|}\right)
   -\frac{|QH|}{\Lambda^2}\left[\ln\!\left(\! 1
   +\frac{\Lambda^2}{m^2}\right)
   \! -\frac{\Lambda^2}{\Lambda^2+m^2}  +2\ln
   \frac{\Gamma\left( \frac{\Lambda^2+m^2}{2|QH|} \right)
   }{\Gamma \left( \frac{m^2}{2|QH|} \right)}\right]\! ,
\end{equation}
where $\psi (x)=\ud\ln\Gamma (x)/\ud x$ is the Euler dilogarithmic 
function. Here the l.h.s. originates from $V_{st}$ and the r.h.s. 
from $V_S$. 

Let us consider f\mbox{}irst the standard case with $\kappa ,\lambda =0$ 
and $h=-m/G$. Then the l.h.s. is a constant $\tau^{-1}=2\pi^2/G\Lambda^2
N_c$. Fig. 1 (left panel) illustrates this pattern. One sees that at 
$H=0$ the system is in the subcritical regime of dynamical symmetry 
breaking. The introduction of a constant magnetic f\mbox{}ield, however 
small it might be, changes radically the dynamical symmetry breaking 
pattern: due to the singular behaviour of the r.h.s. of Eq. (\ref{gap}) 
close to the origin the curves corresponding to the r.h.s. and l.h.s. 
will always intersect and the value of $m$ where this happens is a 
minimum of $V(m)$. One concludes that in the theory with just 
four-fermion interactions the ef\mbox{}fective potential has only one 
minimum at $m>0$, and this property does not depend on the strength of 
the f\mbox{}ield $H$. 

In the theory with four-, six-, and eight-quark interactions one can 
f\mbox{}ind either one or two local minima at $m>0$. We illustrate these 
two cases in the central panel of fig. 1. Namely, the upper full curve 
$f$ (r.h.s. of Eq. (\ref{gap}) for $|QH|\Lambda^{-2}=0.5$) has only one 
intersection point with the bell-shaped curve $u$ (l.h.s. of Eq. 
(\ref{gap}) for $G\Lambda^2=3,\,\kappa\Lambda^5=-10^3,\,\lambda
\Lambda^8=3670$). This point corresponds to a single vacuum state of the 
theory. The other full curve $f$ for $|QH|\Lambda^{-2}=0.1$ has three 
intersections with the same curve $u$. These intersections, successively, 
correspond to a local minimum, a local maximum and a further local 
minimum of the potential. The f\mbox{}irst minimum catalyzed by a 
constant magnetic f\mbox{}ield (that is, a slowly varying f\mbox{}ield) 
is then smoothed out with increasing $H$. It ceases to exist at some 
critical value of $|QH|\Lambda^{-2}$, from which on only the large 
$M_{\dyn}$ solution survives. This is shown in the right panel of 
f\mbox{}ig. 1. The new phenomenon might be a clear signature of 
eight-quark interactions. 

\begin{figure}[t]
\hspace{0cm}{\includegraphics[height=5.6cm]{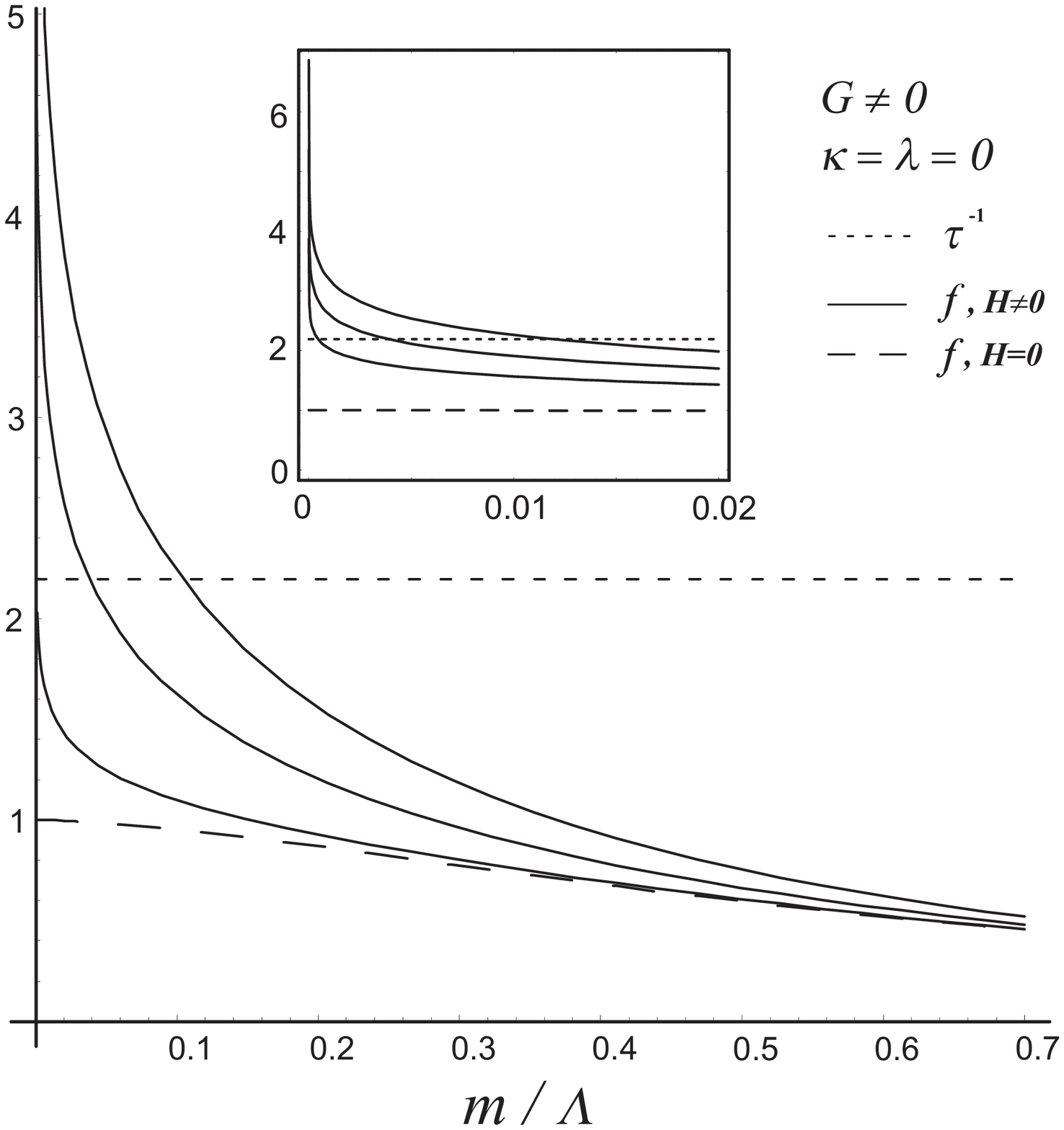}}  
\hspace{0.1cm}{\includegraphics[height=5.6cm]{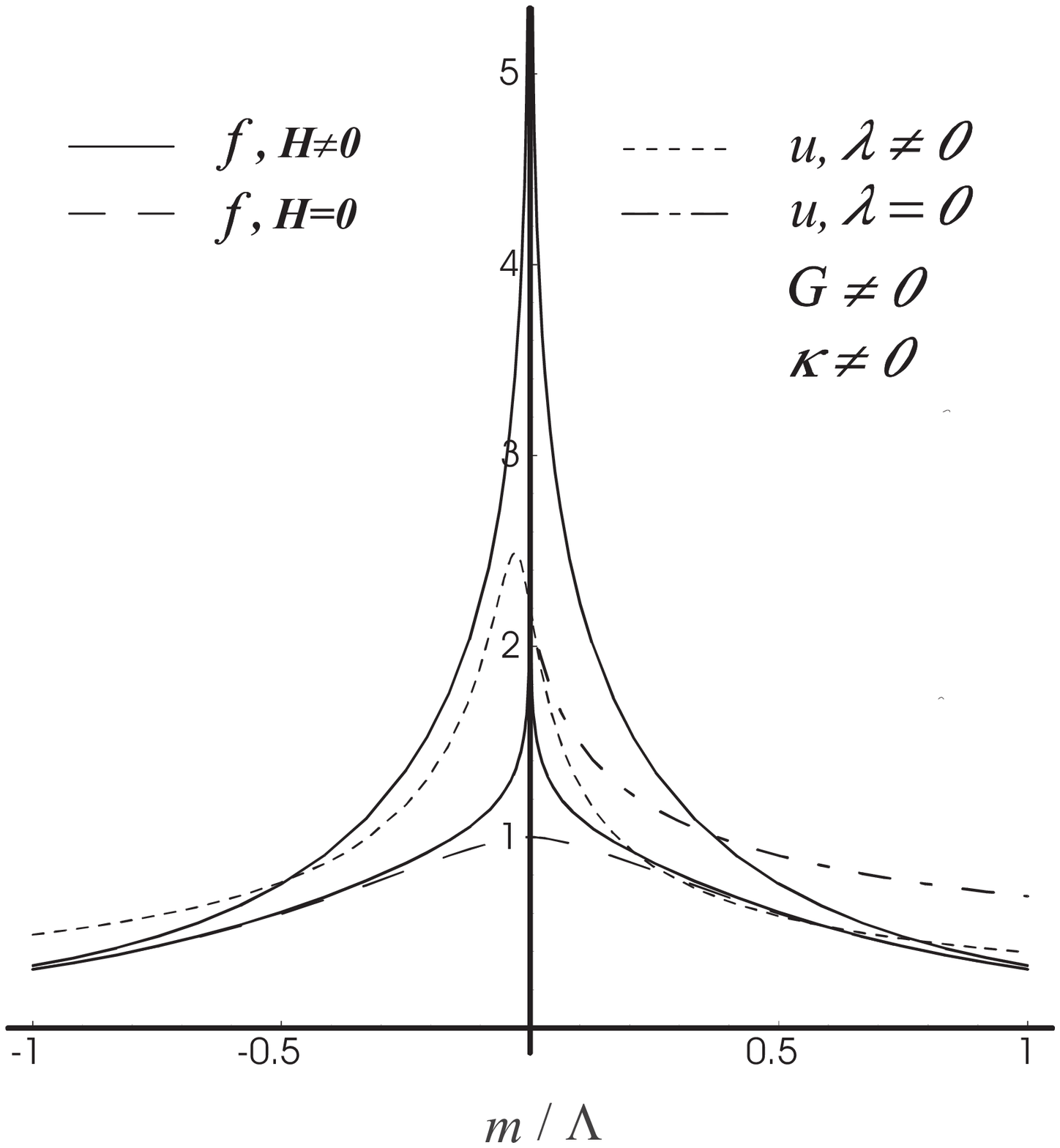}}  
\hspace{0.1cm}{\includegraphics[width=5.0cm]{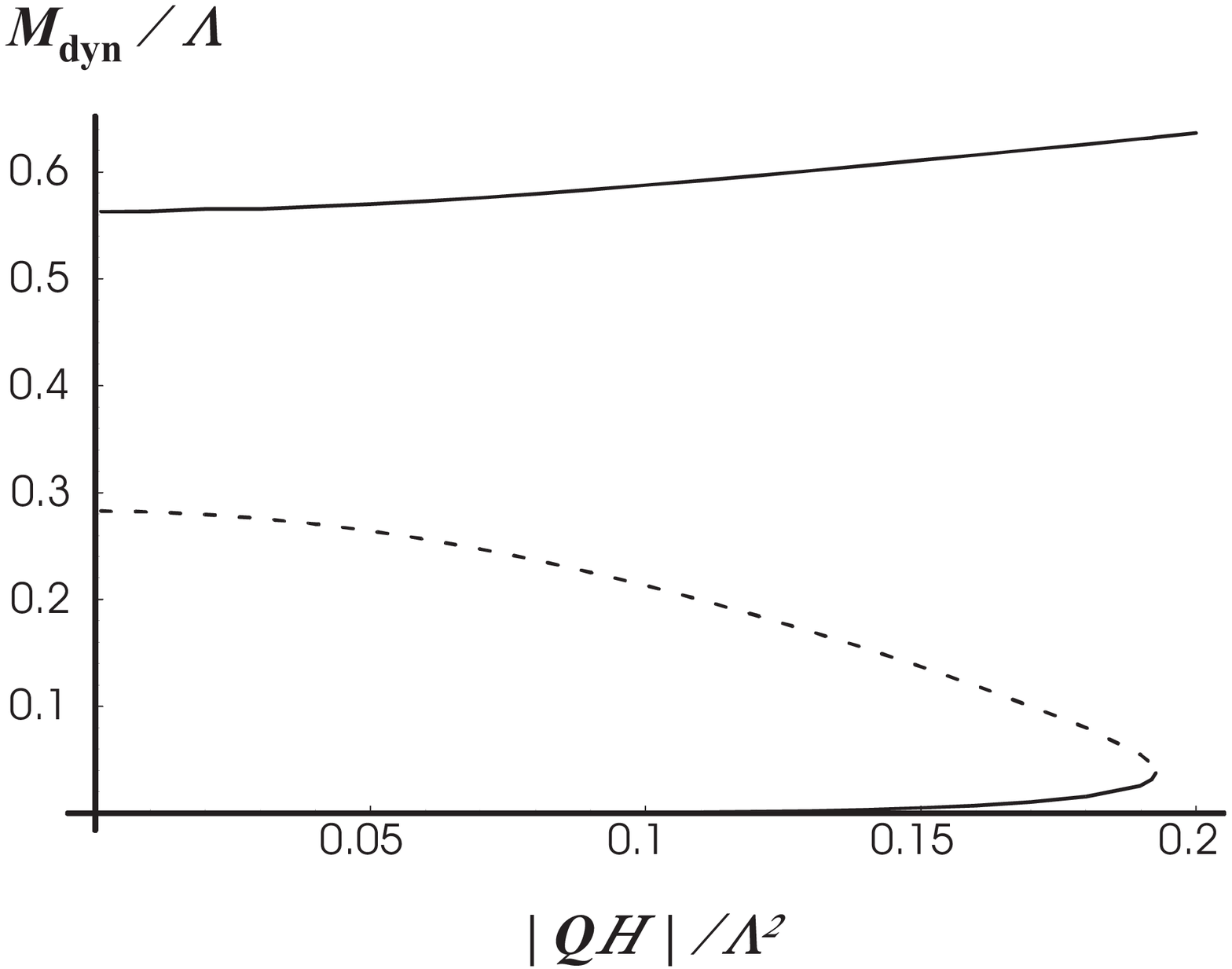}}  
\caption{\underline{Left:} The l.h.s. (straight short-dashed line) and 
  the r.h.s. of Eq. (\ref{gap}) at $\kappa,\lambda =0$ and 
  $G\Lambda^2=3$ as functions of $m/\Lambda$ for four dif\mbox{}ferent 
  values of $H$: full curves (top to bottom) correspond to $|QH|
  \Lambda^{-2}=0.5;\,0.3;\,0.1$, and the dashed curve to $H=0$. Box 
  insert: close-up of region around origin with solid lines for $|QH|
  \Lambda^{-2}=0.2;\,0.15;\,0.1$. 
  \underline{Centre:} 
  The l.h.s. (short-dashed line) and the r.h.s. of Eq. (\ref{gap}) at 
  $\kappa\Lambda^5=-10^3\, ,\lambda \Lambda^8= 3670$ (or $\lambda=0$), 
  and $|QH|\Lambda^{-2}=0.5;\,0.1;\, 0$. 
  \underline{Right:} The dimensionless dynamical mass $M_\dyn/\Lambda$
  as a function of the dimensionless magnetic f\mbox{}ield
  $|QH|\Lambda^{-2}$. The full lines are minima, the dashed line 
  maxima. Up to $|QH|\Lambda^{-2}=0.084$ the smaller $M_\dyn/\Lambda$ 
  correponds to the deeper minimum of the potential; from this value 
  on the larger solution becomes the stable configuration.} 
\end{figure}

This work has been supported in part by grants provided by FCT: 
POCI/2010, FEDER, POCI/FP/63930/2005 and POCI/FP/81926/2007.

\end{document}